\documentclass[prb,a4paper,twocolumn,showpacs,floatfix]{revtex4-1}
	
    \usepackage[T1]{fontenc}

    \usepackage{amssymb}
    \usepackage{amsmath}
    \usepackage{amsbsy}
    \usepackage{bm}
    \usepackage{graphicx}
    \usepackage[bookmarksnumbered,pdfpagelabels=true,plainpages=false,colorlinks=true,linkcolor=blue,citecolor=red,urlcolor=blue]{hyperref}
 \usepackage{xcolor}
 \usepackage{calrsfs}
\DeclareMathAlphabet{\pazocal}{OMS}{zplm}{m}{n}

\begin{document}
\title{Soliton like magnetization textures in noncollinear antiferromagnets}
\author{Camilo Ulloa\email{camilo.ulloa.o@gmail.com},  A.S. Nunez\email{alnunez@dfi.uchile.cl}}%

\affiliation{Departamento de F\'isica, Facultad de Ciencias F\'isicas y 
Matem\'aticas, Universidad de Chile, Casilla 487-3, Santiago, Chile}

\date{\today}

\begin{abstract}
We show that proper control of magnetization textures can be achieved in non-collinear antiferromagnets. This opens the versatile toolbox of domain wall manipulation in the context of a different family of materials. In this way, we show that non-collinear antiferromagnets are a good prospect for  applications in the context of antiferromagnetic spintronics. As in many non-collinear antiferromagnets the order parameter field takes values in SO(3). By performing a gradient expansion in the energy functional we derive an effective theory that accounts for the physics of the magnetization of long wavelength excitations. We apply our formalism to static and dynamic textures such as domain walls and localized oscillations, and identify topologically protected textures that are spatially localized. Our results are applicable to the exchange-bias materials Mn$_3$X, with X= Ir, Rh, Pt.
\end{abstract}

\maketitle

\section{Introduction}
In antiferromagnetic materials the exchange coupling among neighboring spins favors antiparallel arrangements. Because of this interaction the system is led to an ordered magnetic state where the magnetization of different sublattices is oriented in a way that the overall magnetization is canceled. This order drives the system into a robust collective behavior with soft modes that can be controlled with the aid of external magnetic fields.
Recently, the notion that spintronic effects analogous to the ones in ferromagnets can be exhibited by antiferromagnetic systems has received attention from the theoretical \cite{MacDonald, Nunez, Gomonay,Nunez2,Troncoso} and experimental\cite{Urazhdin} viewpoints. The advantages of antiferromagnetic systems come due to a variety of reasons. For example they do not display stray fields, they display high frequency response (in the terahertz range), and finally the fact that antiferromagnetism is observed more often  and at much softer conditions than ferromagnetism. 

A promising development in the context of antiferromagnetic spintronics is the fact that it is possible to engineer magnetic textures, such as domain walls (DWs), in antiferromagnetic systems \cite{Logan}. The problem of domain-wall manipulation in antiferromagnetic systems has been studied in some recent papers\cite{Hals} where, using the collective coordinates approach,  it was shown that the domain wall center obeys Newton's law of motion. This opens the possibility of implementing domain-wall control over antiferromagnets in the same fashion as it is done in ferromagnets. In ferromagnetic systems magnetization textures, smooth modulations in the magnetization field, can be controlled in a diversity of manners, for example through the action of external fields or currents. Research in the field of magnetic domain wall manipulation has been growing steadily\cite{ReviewDWmotion}. The driving force behind this research is the potential applications in the context of information technologies. An example of these applications, is the racetrack\cite{racetrack} configuration where domain walls are driven across a ferromagnetic wire by a current. Domain wall manipulation has also been shown as an alternative to electronic logic circuits\cite{Omari}.  

In this paper we propose that magnetic textures can also be found and controlled in non-collinear antiferromagnets, that is, antiferromagnets whose underlying magnetic sub-lattices are not oriented along the same magnetic axis. Our main result is the theoretical characterization of the dynamics of domain walls in a non-collinear antiferromagnet. 
While our qualitative results apply to a wide family of non-collinear antiferromagnets, we will focus our attention on the magnetic degrees of freedom Mn$_3$Ir. This material has been studied extensively due to its importance as the pinning agent in exchange bias controlled spin-valves devices.
Mn$_3$Ir is regarded as a crystal with fcc structure with Mn atoms lying of the centers of the faces of each cube. The Mn sublattices are two-dimensional kagome lattices lying in the planes perpendicular to the (111) direction. Due to the frustration within each triangular plaquette, isolated isotropic kagome lattices are known examples of disordered spin systems \cite{disorder}. On the contrary the Mn spins in Mn$_3$Ir display a quite strong three-sublattice triangular (T1) magnetic order up to a transition temperature of $\sim 950\;K$\cite{order}. The stability of magnetic order is due primarily to the exchange interaction among the kagome planes and to anisotropy \cite{LeBlanc, Szunyogh,Hemmati}.
Following \cite{Chen}, we use as a minimal model for the physics of the magnetization in Mn$_3$Ir we start with a single nearest neighbour antiferromagnetically coupled kagome lattice of classical spins with appropriately tuned anisotropy terms. 

\section{Basic Model} 
\begin{figure}[htbp] 
   \centering
   \includegraphics[width=0.5\textwidth]{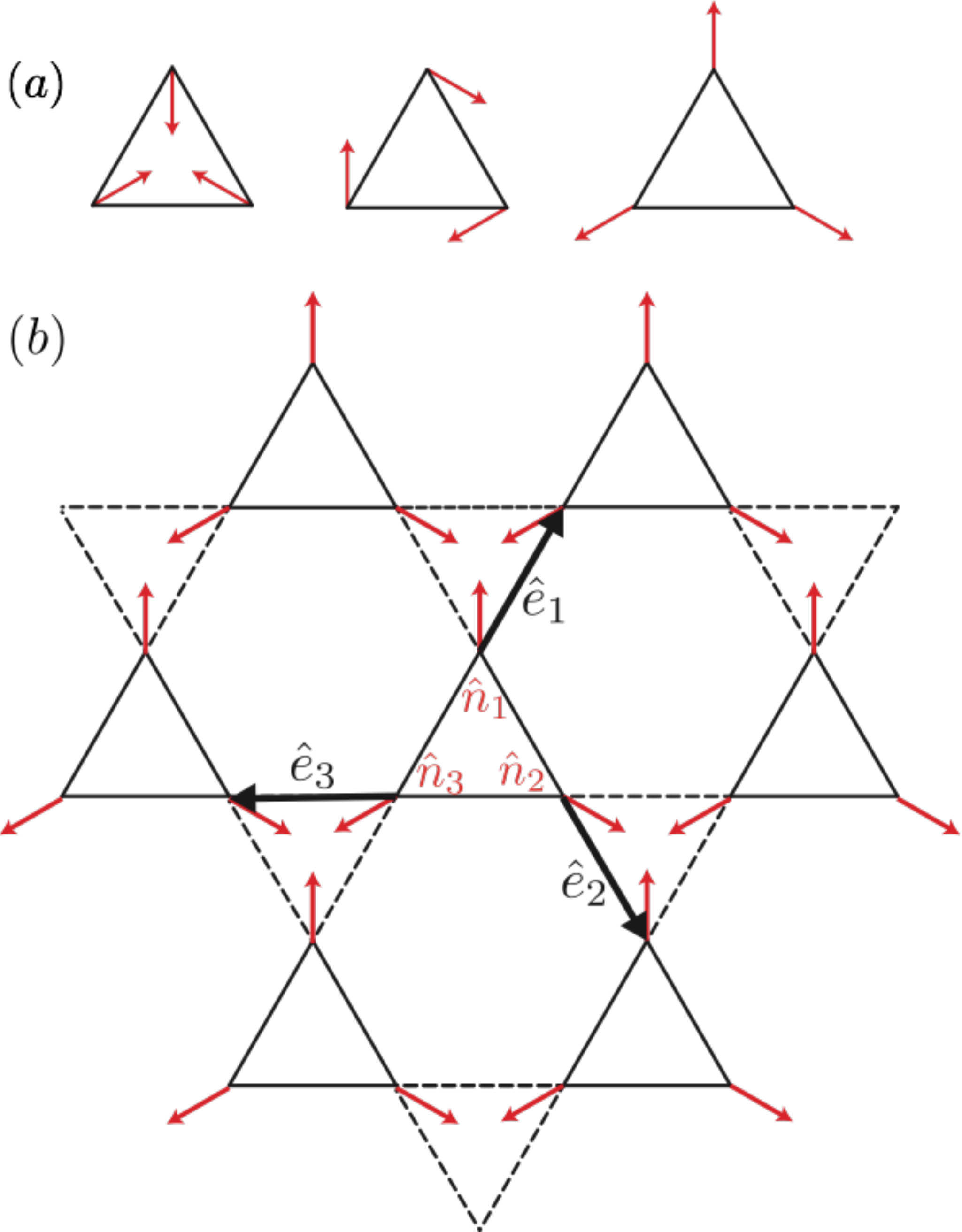}
   \caption{(a) Different configurations of a given triangle achieved through different rotations around the out of plane axes. An arbitrary configuration is encoded by a smooth distributions of such rotations. (b) Kagome lattice in the (111) plane in $Mn_3Ir$ where $Mn$ atoms are at each corner of a basis triangle. The basis  vectors ${\bf n}_1 = (0, 1, 0)$, ${\bf n}_2 = (\sqrt{3}/2,-1/2,0)$ and ${\bf n}_3=(-\sqrt{3}/2,-1/2, 0)$ defined at every point in the lattice are shown. The vectors ${\bf e}_i$ point towards the nearest neighbours of each site. These vectors are used in the gradient expansion in the continuum approximation and are defined as ${\bf e}_1 = (\cos \pi/3, \sin \pi/3, 0)$, ${\bf e}_2 = (\cos \pi/3,-\sin \pi/3, 0)$, and  ${\bf e}_3 = (-1, 0, 0)$.}
   \label{fig:1}
\end{figure}

The minimal model for magnetization dynamics of the Mn atoms in a (111) plane of Mn$_3$Ir  starts from a system of classical spins located at the vertices of a kagome lattice. These spins correspond to the magnetic degrees of freedom of the planes perpendicular to the (111) direction. The  Hamiltonian of the spin system contains two main contributions. On one side we have the exchange interaction, characterized by an exchange constant $J$, between nearest neighbors that favor antiparallel arrangements. On the other we have a strong anisotropy energy that favors orientation in the axis towards the center of the triangles, this energy is characterized by an anisotropy constant $K$, and an anisotropy that penalizes the out-of-plane orientations characterized by $K_z$. The resulting Hamiltonian becomes:  
\begin{equation}
\mathcal{H}=J\sum_{\langle \mathbf{r},\mathbf{r}^\prime \rangle} \mathbf{S}_{\mathbf{r}}\cdot \mathbf{S}_{\mathbf{r}^\prime}-K\sum_{\mathbf{r}}\left(\mathbf{n}_{\mathbf{r}}\cdot\mathbf{S}_{\mathbf{r}}\right)^2 + K_z\sum_{\mathbf{r}}\left(\hat{\mathbf{z}}\cdot\mathbf{S}_{\mathbf{r}}\right)^2,
\label{eq: Hamiltonian}
\end{equation}
where the anisotropy axis, ${\bf n}_{\bf r} $, are defined on each triangular element of the kagome lattice as illustrated on Fig.(\ref{fig:1}), and $\hat{\mathbf{z}}$ is the perpendicular axis to the plane. 
As the antiferromagnetic coupling favors configurations where all the moments in each triangle cancels one other, then if we consider any solid rotation of the moments in one triangle this condition still. 
Following this idea we start from a given minimum (say, all spins pointing outward) and parameterize the configuration on any point in the lattice by a rotation matrix\cite{Dombre,Andreev}:
$
\mathbf{S}_{\mathbf{r}}=\mathcal{R}(\mathbf{r})\lbrace\mathbf{n}_\mathbf{r}+a\left[\mathbf{L}-(\mathbf{L}\cdot \mathbf{n}_\mathbf{r})\mathbf{n}_\mathbf{r} \right]\rbrace,
$
where $\mathbf{L}$ is the canting field assumed to be small.
Restricting the description to the low energy structures we can assume the behavior of the rotation matrix $\mathcal{R}$ to be smooth, varying only across long length scales.
Following \cite{Dombre} we write the Lagrangian of the system in terms of the variables $\mathbf{L}$ and $\mathcal{R}$ and express it within the smooth gradient approximation. We then proceed to solve the Euler-Lagrange equations for the field $\mathbf{L}$ finding:
\begin{equation*}
T\mathbf{L}=\mathcal{R}^{-1}\partial_{t}\mathcal{R}, 
\end{equation*}
where  $T_{\alpha\beta}=\delta_{\alpha\beta} - \frac{1}{3}\sum_{i}n_{i\alpha}n_{i\beta}$. Replacing this solution into the action we are led to an effective Lagrangian density involving only the $\mathcal{R}$ field:
\begin{equation*}
L=-\dfrac{\hbar^2}{2\sqrt{3}Ja^2}\text{Tr}[(\mathcal{R}^{-1}\partial_{t}\mathcal{R})^2] - \mathcal{E}_{ex}(\mathcal{R})-\mathcal{E}_{ani}(\mathcal{R}).
\end{equation*}
The anisotropy coupling favors two configurations, either all spins point toward the center of each triangle or away from it. This state of affairs leave us with two ground states that are degenerate and the main discussion that follow concerns mainly with magnetic textures that connect those states smoothly.
In particular we focus on states that can be obtained from the uniform ground state by a smoothly varying rotation.
It is a straightforward calculation to show that a gradient approximation of the exchange energy functional give us:
$$
\mathcal{E}_{ex}[\mathcal{R}]=\frac{Ja^2}{2}{\rm tr}\left(g^{ij}\mathcal{L}_i\;\mathcal{L}_j\right)
$$
where $\mathcal{L}_i=\mathcal{R}^{-1}\partial_i\mathcal{R}$
and $g^{ij}_{\alpha\beta}=e^i_1 e^j_1 n^{3}_\alpha n^{2}_\beta+e^i_2 e^j_2 n^{1}_\alpha n^{3}_\beta+e^i_3 e^j_3 n^{2}_\alpha n^{1}_\beta $. In the last expression the vectors $\mathbf{e}_i$ correspond to the ones defined in Fig. (\ref{fig:1}b). The anisotropy contribution to the energy is:
$$
\mathcal{E}_{ani}[\mathcal{R}]=-K\sum_i\left(n^i\cdot\mathcal{R}n^i \right)^2+K_z\sum_i\left(\hat{z}\cdot\mathcal{R}n^i \right)^2.
$$

\section{Spin wave spectra} 
\begin{figure}[htbp] 
   \centering
   \includegraphics[width=0.50\textwidth]{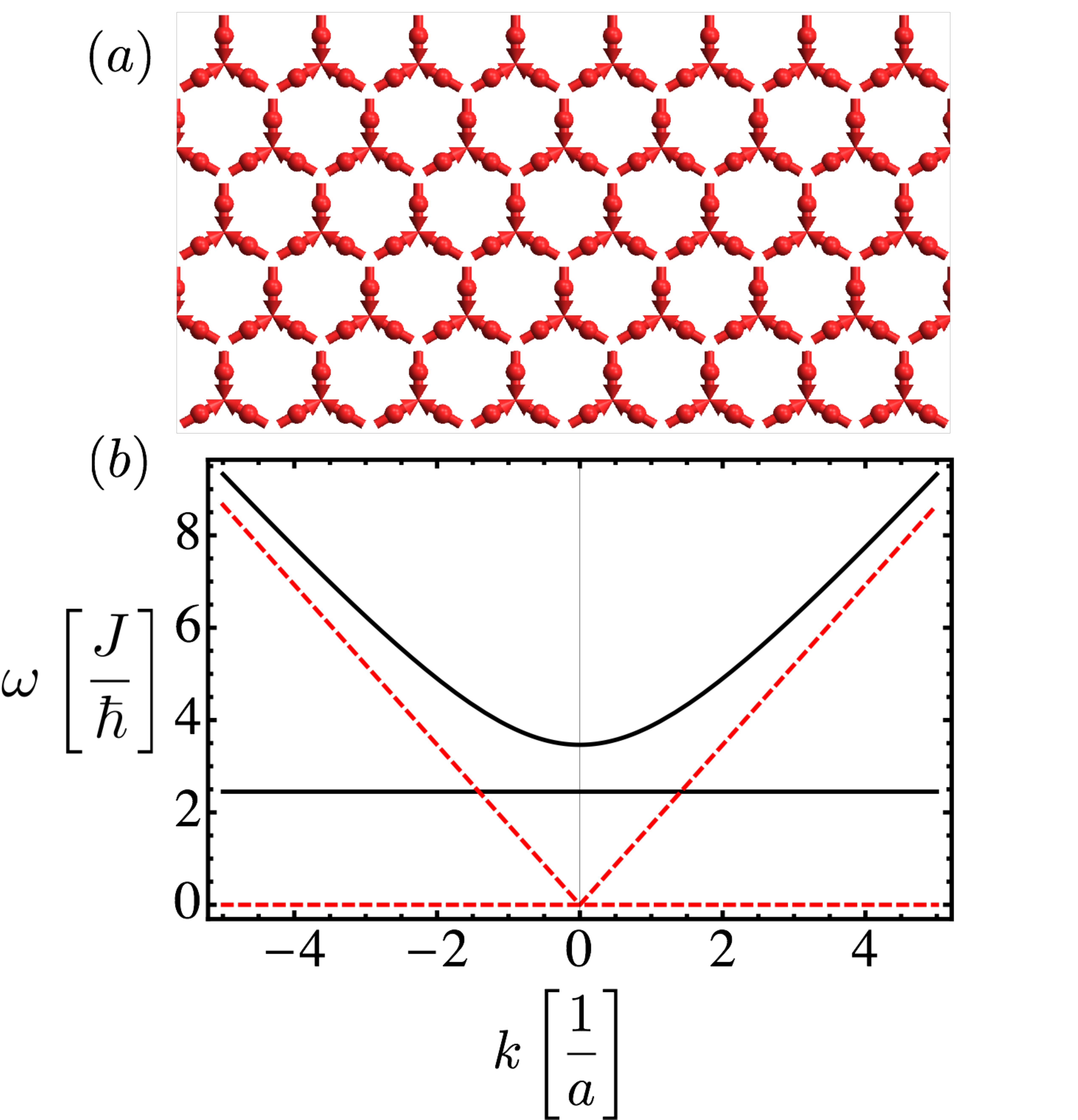}

   \caption{(a) Homogenous state with all the spins pointing toward the center of the triangles. This state is degenerated with the state with all the spins pointing away from the center of each triangle. (b) Dispersion relations for the spin wave spectrum of the homogeneous state. Solid lines correspond to the case with anisotropy and describe two branches one being a flat band with zero group velocity and another with a Klein-Gordon-like dispersion. Dashed line represent the dispersion relations of the isotropic case.}
   \label{fig:2}
\end{figure}

Now we proceed to analyze the spin wave spectra of the system around the ordered phase (Fig.\ref{fig:2}(a)). Here we call this phase just as homogeneous phase. To achieve this goal we describe the state of the system assuming that $\mathcal{R}$ is a rotation matrix made of Euler angles, $\mathcal{R}(\phi,\theta,\psi)=\mathcal{R}_{Z'}(\psi)\mathcal{R}_{X}(\theta)\mathcal{R}_Z(\phi)$. Calculating perturbations around the homogeneous phase we derive the effective action for the spin waves. As the perturbations are around the identity matrix, we found that the variables $\phi$ and $ \psi$ correspond to the same rotation. Defining $ \chi =\phi+\psi$ the equations of motion are:
\begin{equation}\label{eq:02}
\partial_{t}^{2} \chi -\dfrac{3a^2J^2}{\hbar^2}\nabla^2  \chi +\dfrac{12KJ}{\hbar^2}\chi =0,
\end{equation}
\begin{equation}\label{eq:03}
\partial_{t}^{2}\theta+\dfrac{6J(K+K_z)}{\hbar^2}\theta=0.
\end{equation}
The spin wave spectra is split in a dispersion-less flat band with frequency $\omega^2=\dfrac{6J(K+K_z)}{\hbar^2}$ independent of the wave-vector and, a Klein-Gordon-like branch with frequency $\omega^2= \dfrac{3a^2J^2}{\hbar^2}k^2+\dfrac{12KJ}{\hbar^2}$, as is shown in Fig.\ref{fig:2}(b). 
In absence of anisotropy the spin wave branches become $\omega=0$ and $\omega=vk$ with $v=\sqrt{3}aJ/\hbar$ in agreement with the results of \cite{Harris,LeBlanc}. 
The presence of the flat band is a direct consequence of the absence of interlayer couplings within our model. If we consider interlayer exchange interaction including the terms outside the plane (111)  in the gradient expansion, the flat band is modified as is shown in \cite{LeBlanc,Gomonaj1}.


\section{Soliton-like structures } 
\begin{figure}[htbp] 
   \centering
   \includegraphics[width=0.5\textwidth]{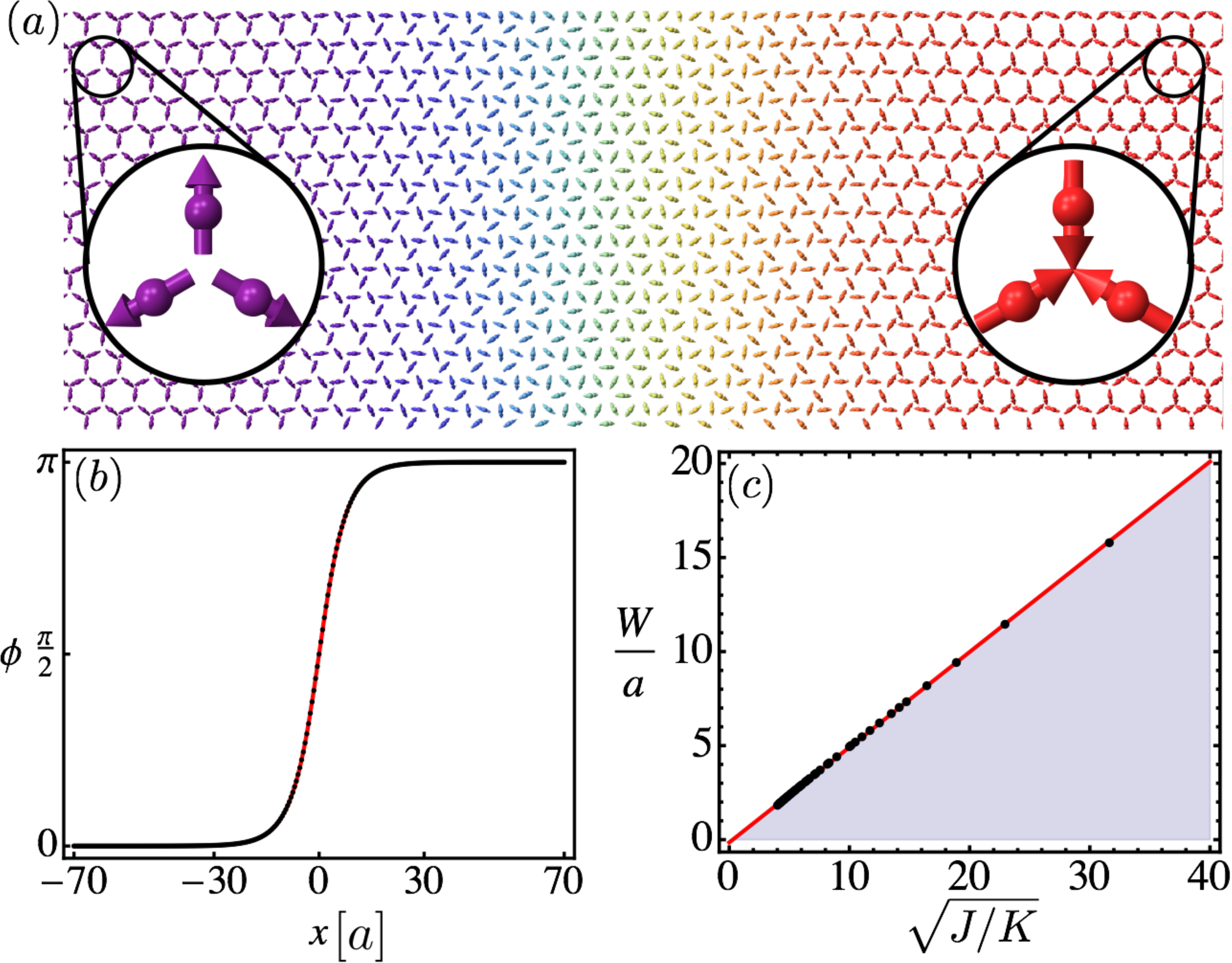} 
   \caption{(a) Typical shape of a domain wall in kagome lattice. (b) Numerical fit of the soliton solution, black dots are the result of numerical simulation, red line corresponds to fitted solution $\phi=2 \tan^{-1}(\exp(x/W))$. (c) Width dependence on anisotropy. Black dots are numerical results while red dashed line is the fitted curve which has a slope equal to 1/2.}
   \label{fig:3}
\end{figure}

We continue our discussion on magnetic textures looking at possible DWs in the order parameter field. We parameterize the rotation at each point by an angle $\phi$ and a rotation axis parallel to the $z$-axis (out of the plane). The Lagrangian density can be expressed in terms of $\phi$:
$$
L=\frac{\hbar^2}{\sqrt{3}Ja^2}(\partial_t \phi)^2 - \sqrt{3}J \left(\nabla\phi\right)^2+\frac{4\sqrt{3}K}{a^2}\cos^2\phi.
$$
This correspond to the well known sine-Gordon model whose equation of motion is:
$$
\partial^2_t\phi-c^2\nabla^2\phi+\frac{m^2c^4}{\hbar^2}\sin(2\phi)=0
$$
where we have defined the spin-wave velocity $c^2=3J^2a^2/\hbar^2$ and the mass parameter $m^2c^4/\hbar^2=6KJ/\hbar^2$.
The solutions of this equation have been extensively studied\cite{rajaraman}. Among its most celebrated solutions we can highlight stationary domain walls (where the angle travel all the way from zero to $\pi$) that are characterized by a domain wall width $W=  a\sqrt{J/K}/2$.  For $Mn_3Ir$ rough estimates of the parameters lead us to $W\sim 1\text{-}10a$ \cite{Szunyogh}. The profile of the domain wall has a soliton-like form: $\phi=2 \tan^{-1}(\exp(x/W))$ (Fig.\ref{fig:3}(b)). To characterize the domain walls of we have solved numerically Landau-Lifshitz-Gilbert equation with an effective field derived from Eq.(\ref{eq: Hamiltonian}). By setting periodic boundary conditions in the exchange field we have enforced a domain wall within our system and let the system to relax. The domain wall profile is then optimized and its width determined by fitting to a soliton like shape with adjustable width. The results are displayed in Fig.(\ref{fig:3}) and are in remarkable agreement with the long wave-length description of the continuum model.

\begin{figure}[htbp] 
   \centering
   \includegraphics[width=0.5\textwidth]{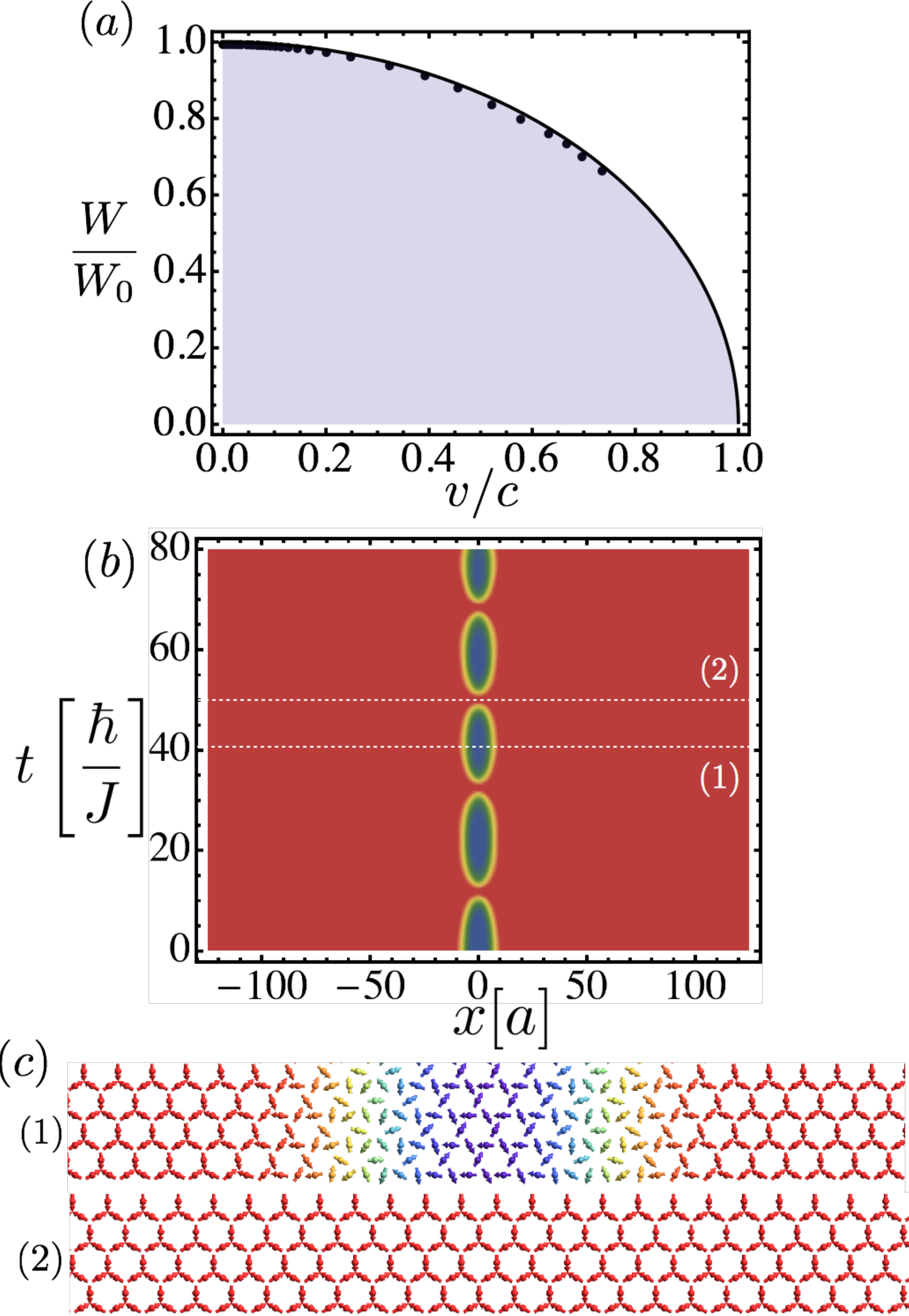}

   \caption{(a) Contraction of the width $W$ of a moving DW as function of speed $v$. The simulations were performed setting the easy axis anisotropy by $K=0.025J$, and the hard axis anisotropy by $K_z=0.1J$. The results of the Landau-Lifshitz equation shows perfect agreement with the Lorentz contraction factor $\sqrt{1-(v/c)^2}$, full line, that can be inferred from the sine-Gordon equation.  (b) Time evolution of the orientiation $\phi$ in the case of a breather state with frequency $\omega=0.25$. The results correspond to the solution of the Landau-Lifshitz equation with the same paremeters as in Fig.\ref{fig:4}(a). The color code is the same as the one used for the DW. (c) Snapshot at time (1) and (2) showing the orientation of the local moments in the texture. }
   \label{fig:4}
\end{figure}

Along with the stationary domain walls just described the sine-Gordon model allows for mobile textures. As it is well known the profile of a domain wall moving with velocity $v$ is contracted by the Lorentz factor leading to a solution $\phi=2 \tan^{-1}(\exp[(x-vt)/W_0\sqrt{1-(v/c)^2}])$. We have verified this behavior using our simulations based on the Landau-Lifshitz equation. The results are shown in Fig.\ref{fig:4}(a).

Among the other localized excitations that are associated with the sine-Gordon equation, we have focused on the stationary breather solution \cite{librosg}, $\phi=2\tan^{-1}\left[\sqrt{1-\omega^2}\cos(\omega t/\tau)/\omega \cosh(x\sqrt{1-\omega^2}/\lambda) \right]$, where $\tau=\hbar/mc^2\sqrt{2}$, and $\lambda=\hbar/mc\sqrt{2}$.  This solution represents a localized oscillation of the orientation of the local moments around the anisotropy axes. The numerical solution of the Landau-Lifshitz equation that is consistent with this state is shown in Fig.\ref{fig:4}(b)-(c). 

\section{Topological defects} 
The topology of the order parameter space ($SO(3)$) opens a variety of possible topologically protected defects \cite{Mermin}. For example, the first homotopy group being $\pi_1(SO(3))=\mathbb{Z}_2$, then we have two kinds of configurations. While textures which belong to class 0 can be continuously deformed to the uniform state, textures which belong to class 1 cannot. The latter are known as disgyrations in the context of $^3He$. The coalesence of two disgyrations generates a vortex-like trivial texture (as consequence of  $1+1=0$). However, disgyrations have an energy that grows with system size then are not localized. 

\begin{figure}[htbp] 
   \centering
   \includegraphics[width=0.5\textwidth]{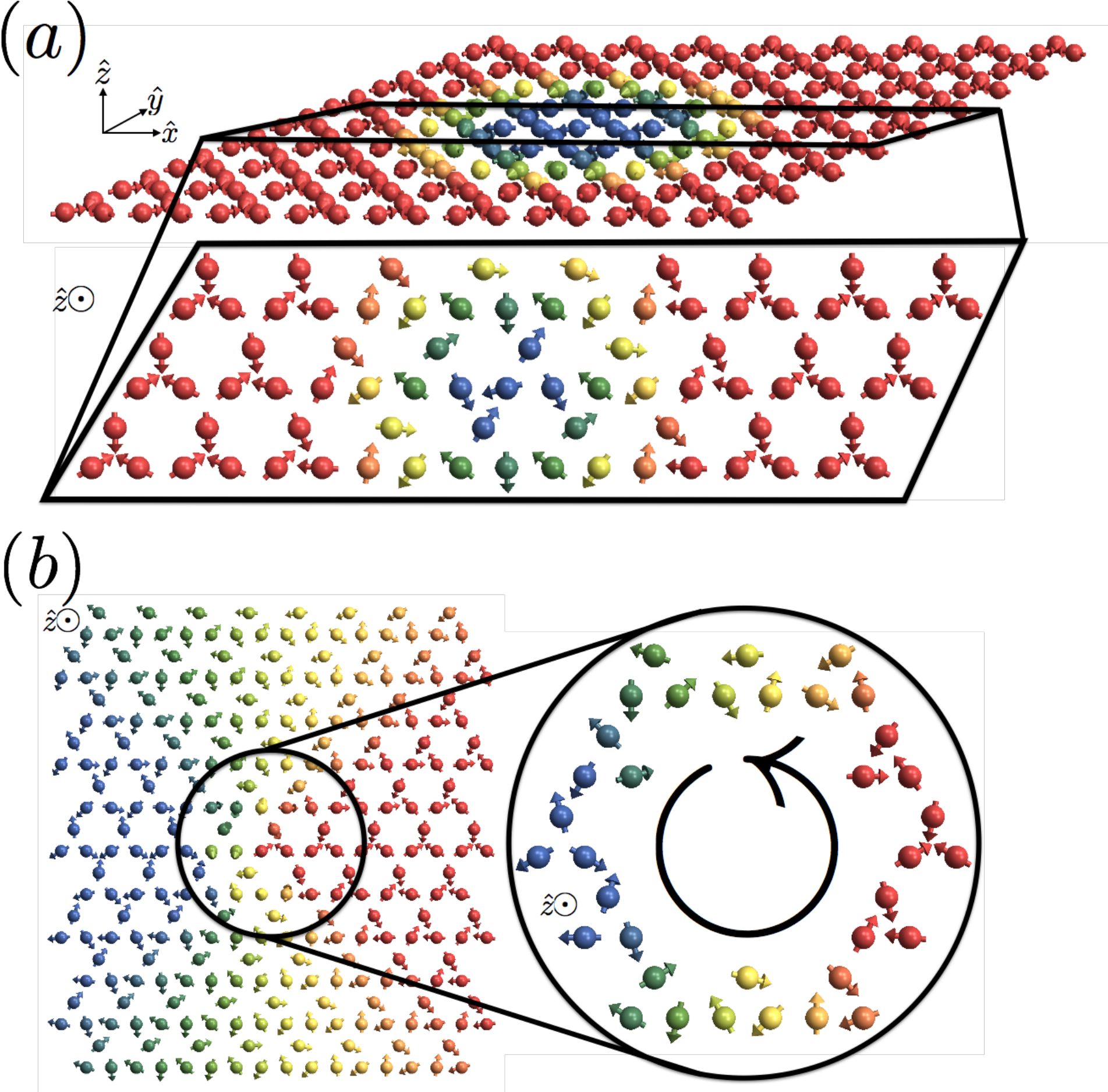}
   \caption{(a)Cartoon of a \emph{lump} texture, given by the parameterization $\mathcal{R}(\eta,\hat{z})\mathbf{n}_\mathbf{r}$, where $\eta(r)= 2\arctan[\exp\ (r-R)/W]$. This is an example of a trivial 2D texture. While this solution is stable, has no topological protection at all so it can be continuously deformed to the homogeneous state.  (b) Cartoon of a class 1 disgyration. This solution is topologically protected because of the non-trivial homotopy $\pi_1(SO(3))$, then is not possible to reach the homogeneous state adiabatically. As this state is not localized, his energy grows with the size of the system. }
\label{fig:5}
\end{figure}
The second homotopy group of $\pi_2(SO(3))$ is trivial and skyrmions are not topologically protected. Nevertheless there are trivial stable solutions as the \emph{lump} solution \cite{librosg}. In Fig.\ref{fig:5} we show examples of textures related with the previous topological properties.

Finally, the third homotopy group of the order parameter space is given by  $\pi_3(SO(3))=\mathbb{Z}$ which opens up the possibility of topologically stable point like defects. Physical realizations of this kind of topological defects have been studied in the context of superfluid $^3$He-A\cite{Nakahara} and topological insulators\cite{Shankar-TI}. 
Our description of a single kagome lattice needs to be extended to include interplane interactions in the gradient expansion. In this case the topological defect is a three-dimensional structure characterized by covering all possible rotations as we move away from its center. The winding number associated with the third homotopy group is given by\cite{Jackiw, Polyakov,Gomonay2}:
$$
Q=\frac{1}{24\pi^2}\int {\rm d}r\; \varepsilon_{\mu\nu\lambda} {\rm tr}\left(\mathcal{L}_{\mu}\mathcal{L}_{\nu}\mathcal{L}_{\lambda}\right)
$$  
where $ \varepsilon$ stands for the fully antisymmetric tensor.
One possible realization of this kind of defects is the Shankar monopole\cite{Shankar}. The idea is to associate to each point of space $\vec{r}=r\hat{n}$ an operator that rotates around the $\hat{n}$-axis an angle $\chi(r)$. If $\chi$ is chose to go all the way from zero at the origin to $2\pi$ away from the monopole we can see that whole parameter space is covered twice. The texture so generated is stable under perturbations and becomes a finite energy topologically protected defect.

\section{Conclusions} In this paper we have addressed the behavior of textures in the order parameter field of non-collinear antiferromagnets. By pursuing a continuum description of the textures we have studied the spin-wave spectra around the homogeneous configuration and the behavior of domain walls.
The spin wave spectra consists of two branches. One correspond to the usual Klein-Gordon-like dispersion relation while the other correspond to a flat band whose frequency is independent of the wave-number. Domain wall structure behave in a similar fashion than Bloch type domain walls in common ferromagnets with a characteristic with scaling with the square root of $J/K$ (exchange interaction compared with anisotropy). They are described by an effective sine-Gordon equation that allows us to predict the existence, along with stationary domain walls, of moving domain walls that travel undistorted across the  system.
We have compared the predictions of our continuum theory 
with the results of exact simulations of the Landau-Lifshitz-Gilbert equation and obtained a complete agreement.

Finally we have discussed the topological defects that are allowed by the topology of the order parameter space.
\begin{acknowledgments}
The authors acknowledge funding from Proyecto Fondecyt numbers 1150072, Proyecto Basal FB0807-CEDENNA, and  Anillo de Ciencia y Tecnonolog\'ia ACT 1117
\end{acknowledgments}

\bibliographystyle{elsarticle-num}

\end{document}